\documentclass[journal,twocolumn,10pt]{IEEEtran}
\usepackage{graphicx}
\usepackage{verbatim}
\usepackage{upgreek}
\usepackage{amssymb,amsmath}
\usepackage{color}
\usepackage{epstopdf}
\usepackage{bm}
\usepackage{cite}
\usepackage{array,color}
\usepackage{algorithm}
\usepackage{algpseudocode}
\usepackage{amsmath}
\usepackage{amsfonts}
\usepackage{amsthm}
\usepackage{graphics}
\usepackage{epsfig}
\usepackage{soul}
\usepackage{booktabs}
\usepackage{multirow}
\soulregister\cite7
\soulregister\ref7
\usepackage{subfigure}
\usepackage{tabularx}
\usepackage{makecell}
\newtheorem{lemma}{Lemma}

\usepackage{etoolbox}
\makeatletter
\patchcmd{\@makecaption}
{\scshape}
{}
{}
{}
\makeatletter
\patchcmd{\@makecaption}
{\\}
{.\ }
{}
{}
\makeatother

\ifCLASSINFOpdf
\else
\fi

\begin{document}

\title{\huge Joint Transmitter Design for Robust Secure Radar-Communication Coexistence Systems}

\author{Peng Liu, Zesong Fei,~\IEEEmembership{Senior~Member,~IEEE}, Xinyi Wang,~\IEEEmembership{Member,~IEEE},\\ Zhong Zheng,~\IEEEmembership{Member,~IEEE}, Xiangnan Li,~\IEEEmembership{Member,~IEEE}, and Jie Xu,~\IEEEmembership{Senior~Member,~IEEE}
	
	\thanks{Peng Liu, Zesong Fei, Xinyi Wang, Zhong Zheng, and Xiangnan Li are with the School of Information and Electronics, Beijing Institute of Technology, Beijing 100081, China. (E-mail:bit\_peng\_liu@163.com, feizesong@bit.edu.cn, bit\_wangxy@163.com, zhong.zheng@bit.edu.cn, 7520220063@bit.edu.cn).}
\thanks{Jie Xu is with the School of Science and Engineering and the Future Network of Intelligence Institute (FNii), The Chinese University of Hong Kong, Shenzhen, Shenzhen 518172, China (E-mail: xujie@cuhk.edu.cn).}

}

\maketitle

\begin{abstract}
This paper investigates the spectrum sharing between a multiple-input single-output (MISO) secure communication system and a multiple-input multiple-output (MIMO) radar system in the presence of one suspicious eavesdropper. We jointly design the radar waveform and communication beamforming vector at the two systems, such that the interference between the base station (BS) and radar is reduced, and the detrimental radar interference to the communication system is enhanced to jam the eavesdropper, thereby increasing secure information transmission performance. In particular, by considering the imperfect channel state information (CSI) for the user and eavesdropper, we maximize the worst-case secrecy rate at the user, while ensuring the detection performance of radar system. To tackle this challenging problem, we propose a two-layer robust cooperative algorithm based on the S-lemma and semidefinite relaxation techniques. Simulation results demonstrate that the proposed algorithm achieves significant secrecy rate gains over the non-robust scheme. Furthermore, we illustrate the trade-off between secrecy rate and detection probability.

\vspace{2ex}
\end{abstract}

\IEEEpeerreviewmaketitle
\section{Introduction}
Integrated  sensing and communication (ISAC) has attracted growing interests to efficiently utilize the  
the scarce spectrum resources to support the sensing and communications of ever-increasing wireless devices. \textcolor{black}{In general, dual-functional radar-communication (DFRC) and radar-communication coexistence (RCC) are two main ISAC paradigms \cite{isacliu}. While DFRC enables simultaneous sensing and communication within the same integrated system, RCC exploits the spectrum sharing between coexisting sensing and communication systems, by properly managing their mutual interference.}

There have been numerous efforts aiming at reducing interference between coexisting communication and radar systems. A null-space projection method was proposed in \cite{coexit3}, which projects the radar transmit beam into the null space of the channel towards  the communication receiver \textcolor{black}{for eliminating the interference.} In \cite{coexit5}, convex optimization techniques were employed to jointly design the radar receive beamformer and communication precoder to mitigate interference. In \cite{wangris}, the reconfigurable intelligent surface (RIS) technology was explored to enable the coexistence of multi-input-multi-output (MIMO) radar and multi-input-single-output (MISO) communication systems.

Recently, information security is becoming increasingly important for the practical implementation of of communication systems. Physical-layer security (PLS) technique exploit the difference of wireless channels between the legitimate and the eavesdropping links to achieve secure information transmission. \textcolor{black}{Recently, transmit beamforming and artificial noise (AN) techniques are widely used to improve the security of DFRC systems \cite{Xv2023}\cite{su2021}. However, how to ensure the physical layer security in RCC systems remains to be studied.} It is worth noting that although radar interference is typically harmful to communication systems, such co-channel interference can be used to jam eavesdroppers\cite{JSTSP2012}. Motivated by \cite{JSTSP2012}, prior work \cite{Liuiccc} designed a PLS scheme by employing the well-designed radar signals to jam the eavesdropper. Furthermore, in \cite{chu2022}, additional AN was introduced to jointly optimize the transmit beams of radar and communication systems, with the aim of suppressing the signal-to-interference-plus-noise ratio (SINR) at the eavesdropper. However, these aforementioned works for RCC system primarily focused on the design from a communication system perspective, without fully analyzing the impact of secrecy transmission schemes on radar performance. Additionally, previous works \cite{su2021}\cite{chu2022} mainly considered  the deterioration of eavesdropper's SINR, rather than directly \textcolor{black}{maximizing the secrecy rate,} which cannot intuitively reflect the secure transmission performance. Moreover, in practice, eavesdroppers are generally non-cooperative, and there thus exist errors in \textcolor{black}{the acquired channel station information (CSI).} Hence,  non-convex form of secrecy rate and the consideration of imperfect CSI make the robust design of secure transmission schemes in RCC system more challenging.

In this paper, we consider the robust \textcolor{black}{joint transmitter design for the} coexistence of MIMO radar and MISO secure communication systems, \textcolor{black}{in which the CSI is subject to bounded errors due to the imperfection in the  acquisition  process.}  Thereafter, the radar waveform and communication beamforming vector are jointly optimized to improve the worst-case secrecy rate of the communication system, \textcolor{black}{subject to constraints on the interference from the communication system to radar and the radar beampattern mismatch level.} \textcolor{black}{To solve the challenging radar-constrained worst-case secrecy rate maximization problem,}  a two-layer robust cooperative algorithm based on the S-lemma and  semidefinite relaxation (SDR) is proposed. Finally, simulation results validate   \textcolor{black}{the performance of the secure communication and radar systems achieved by} the proposed algorithm, revealing the trade-off between radar detection and communication secrecy rate.

\section{System Model}

We consider a coexistence system of downlink communication and collocated MIMO radar operating in the same frequency band. The base station (BS) equipped with a transmit uniform linear array (ULA) of $N$ antennas serves a single-antenna user (Bob) in the presence of an eavesdropper (Eve) attempting to overhear the transmission. Meanwhile, the MIMO radar with $M_t$-element transmit ULA and $M_r$-element receive ULA is attempting to detect a point target in the far field. The transmit and receive steering vectors of radar antenna array with angle $\theta$ is expressed as $\mathbf{a}_t(\theta) =\mathbf{a}_r(\theta)= [1, {\rm e}^{j2\pi d \sin(\theta)/\lambda}, \cdots, {\rm e}^{j2\pi (N -1) d \sin(\theta)/\lambda}]^T$, where $\lambda$ denotes the carrier wavelength and $d=\lambda/2$ is the spacing between adjacent antenna elements. 

In practice, obtaining perfect CSI for both legitimate user and eavesdropper is challenging due to the the non-cooperative nature of Eve and the presence of channel estimation and feedback errors. We denote the channels from BS to Bob and Eve as $\mathbf{h}_b \in \mathbb{C}^{N \times 1}$ and $\mathbf{h}_e \in \mathbb{C}^{N \times 1}$, respectively. The interference channels from MIMO radar transmitter to Bob and Eve, and that from BS to the MIMO radar receiver, are represented by $\mathbf{f}_b \in \mathbb{C}^{M_t \times 1}$, $\mathbf{f}_e \in \mathbb{C}^{M_t \times 1}$, and $\mathbf G =[\mathbf{g}_1,\dots,\mathbf{g}_M] \in \mathbb{C}^{N \times M_r}$, respectively. We assume that the CSI errors are norm-bounded by certain given constants \cite{tsp09}. The norm-bounded error model to characterize the imperfect CSI is expressed as
\begin{equation}\small
\begin{aligned}\label{erro}
&\mathcal{H}_b = \{\bm{\Delta_{hb}}\vert\bm{\Delta_{hb}} = \mathbf{h}_b - \bar{\mathbf{h}}_b, \Vert \bm{\Delta_{hb}} \Vert^2 \leq \epsilon_{hb}^2 \}, \\
&\mathcal{H}_e = \{\bm{\Delta_{he}} \vert \bm{\Delta_{he}} = \mathbf{h}_e - \bar{\mathbf{h}}_e, \Vert \bm{\Delta_{he}} \Vert^2 \leq \epsilon_{he}^2 \}, \\
&\mathcal{F}_b = \{\bm{\Delta_{fb}}\vert \bm{\Delta_{fb}} = \mathbf{f}_b - \bar{\mathbf{f}}_b, \Vert \bm{\Delta_{fb}} \Vert^2 \leq \epsilon_{fb}^2 \},\\
&\mathcal{F}_e = \{\bm{\Delta_{fe}}  \vert \bm{\Delta_{fe}} = \mathbf{f}_e - \bar{\mathbf{f}}_e, \Vert \bm{\Delta_{fe}} \Vert^2 \leq \epsilon_{fe}^2 \},\\
&\mathcal{G}_m = \{\bm{\Delta_{m}} \vert \bm{\Delta_{m}} =\mathbf{g}_m -\bar{\mathbf{g}}_m, \Vert \bm{\Delta_{m}} \Vert^2 \leq \epsilon_{gm}^2 \},
\end{aligned}
\end{equation}
where $\bar{\mathbf{h}}_b,\bar{\mathbf{h}}_e,\bar{\mathbf{f}}_b,\bar{\mathbf{f}}_e$, and $\bar{\mathbf{g}}_m$ denote the estimated channel vectors of ${\mathbf{h}}_b,{\mathbf{h}}_e,{\mathbf{f}}_b,{\mathbf{f}}_e$, and ${\mathbf{g}}_m$, respectively, and  $\bm{\Delta_{hb}},\bm{\Delta_{he}},\bm{\Delta_{fb}},\bm{\Delta_{fe}}$, and $\bm{\Delta_m}$ represent the corresponding channel error vectors, with $\epsilon_{hb},\epsilon_{he},\epsilon_{fb},\epsilon_{fe}$, and $\epsilon_{gm}$ denoting the channel error bounds.  \textcolor{black}{We assume that the BS and radar operate in a collaborative way, with the CSI shared with each other.} The received signals at Bob and Eve are respectively expressed as
\begin{align}\small
y^c_b = \mathbf{h}_b^H \mathbf{t} s + \mathbf{f}_b^H \mathbf{x} +z_b, \\
y^c_e = \mathbf{h}_e^H \mathbf{t} s + \mathbf{f}_e^H \mathbf{x} +z_e,
\end{align}
where $\mathbf{t} \in \mathbb{C}^{N \times 1}$ is the communication beamforming vector, and $s$ and $\mathbf{x} \in \mathbb{C}^{M_t \times 1}$ represent the commnunication symbol and radar waveform, respectively. Furthermore, $z_b \sim \mathcal{CN}(0, \sigma_b^2)$ and $z_e \sim \mathcal{CN}(0, \sigma_e^2)$ denote the noise at Bob and Eve, respectively. We denote the covariance matrix of the MIMO radar signal as $\mathbb{E}[\mathbf{x} \mathbf{x}^H]=\mathbf{R}_x \succeq 0 $. The SINRs at Bob and Eve are respectively expressed as
\begin{align}\small
\gamma_b = \frac{\mathbf{h}_b^H \mathbf{t}\mathbf{t}^H\mathbf{h}_b} {\mathbf{f}_b^H\mathbb{E}[\mathbf{x}\mathbf{x}^H]\mathbf{f}_b + \sigma_b^2} = \frac{\mathbf{h}_b^H \mathbf{T}\mathbf{h}_b} {\mathbf{f}_b^H\mathbf{R}_x\mathbf{f}_b + \sigma_b^2} , \\
\gamma_e = \frac{\mathbf{h}_e^H \mathbf{t}\mathbf{t}^H\mathbf{h}_e} {\mathbf{f}_e^H\mathbb{E}[\mathbf{x}\mathbf{x}^H]\mathbf{f}_e + \sigma_e^2} = \frac{\mathbf{h}_e^H  \mathbf{T}\mathbf{h}_e} {\mathbf{f}_e^H\mathbf{R}_x\mathbf{f}_e + \sigma_e^2} ,
\end{align}
where $\mathbf{T} = \mathbf{t} \mathbf{t}^H \succeq 0 $ with rank($\mathbf{T}$) = 1.  \textcolor{black}{By taking into account the CSI errors, the worst-case secrecy rate} of the communication system can be defined as \textcolor{black}{\cite{tsp09}}
\begin{equation}\small \label{secrecycapacity}
\begin{aligned}
C_s = \min \limits _{\substack{\bm{\Delta}_{hb}\in \mathcal{H}_b,\bm{\Delta}_{he}\in \mathcal{H}_e,\\\bm{\Delta}_{fb}\in \mathcal{F}_b,\bm{\Delta}_{fe}\in \mathcal{F}_e,\bm{{\Delta}_{m}}\in \mathcal{G}}} \left[ R_{b} - R_e \right]^+,
\end{aligned}
\end{equation}
where $R_b = \log_2(1+\gamma^c_b)$, $R_e = \log_2(1+ \gamma^c_e)$, and $[\cdot]^+$ = max$\{\cdot, 0\}$.

Considering the echoes in a single range-Doppler bin of the radar detector, the discrete signal vector $\mathbf{y}_r$ received by radar is given as 
\begin{equation}\small\label{radar_signal}
\mathbf{y}_r = \alpha\mathbf{a}_r(\theta) \mathbf{a}_t^T(\theta) \mathbf{x} + \mathbf G^H \mathbf{t} s + \bm z_r,
\end{equation}
where $\theta$ denotes the azimuth angle of the target, $\alpha$ is the complex target coefficient, and $\bm z_r \sim \mathcal{CN}(\bm 0, \sigma_r^2 \mathbf{I})$ denotes the noise vector at the radar receiver, with $\sigma_r^2$ denoting the noise power. The transmit beampattern of radar waveform at given angle $\theta$ is given as
\begin{equation}\small
\begin{aligned}
P_{b}(\theta) &=\mathbf{a}^H({\theta})\mathbf{R}_x\mathbf{a}({\theta}).
\end{aligned}
\end{equation}

The interference from BS to the $m$-th antenna of the radar is
\begin{equation}\small \label{interference}
{w}_m = \mathbf{g}_m^T\mathbf{t}s,
\end{equation}
to evaluate the interference level from BS to radar, we define the Interference-to-Noise Ratio (INR) at the $m$-th antenna of radar as
\begin{equation}\small \label{interference1}
\text{INR}_m = \frac{\mathbb{E}({\vert{w_m}\vert^2})}{\sigma_r^2}=\frac{\text{Tr}(\mathbf{g}_m^*\mathbf{g}_m^T\mathbf{T})}{\sigma_r^2}.
\end{equation}

It should be noted that INR \textcolor{black}{is closely related to} the detection probability of the radar. Under the Neyman-Pearson criterion and by using the generalized likelihood ratio test (GLRT), the asymptotic radar detection probability $P_D$ under a pre-designed false alarm probability $P_{FA}$ is represented as \cite{tsp2006}
\begin{equation}\small \label{detec_prob}
{P_{D}} = 1 - {\mathfrak {F}_{\mathcal {X}_{2}^{2}\left ({\rho }\right)}}\left ({{\mathfrak {F}_{\mathcal {X}_{2}^{2}\left ({\rho }\right)}^{ - 1}\left ({{1 - {P_{FA}}} }\right)} }\right),
\end{equation}
where ${\mathfrak {F}_{\mathcal {X}_{2}^{2}\left ({\rho }\right)}}$ denotes the non-central chi-quare cumulative distribution function (CDF) with 2 degrees of freedom, and ${\mathfrak {F}^{-1}_{\mathcal {X}_{2}^{2}\left ({\rho }\right)}}$ denotes its inverse function. The non-centrality parameter $\rho$ is given by \cite{1998dec} 
\begin{equation}\small\label{detec_prob2}
\rho = {|\alpha | ^{2}}L\text{Tr} \left ({{{\mathbf {A}}\mathbf{R}_x {{\mathbf {A}}^{H}} {{\left ({{{{\mathbf {G}}^{T}}\mathbf {T}{{\mathbf {G}}^{*}} + \sigma _{r}^{2}{\mathbf {I}}} }\right)}^{ - 1}}} }\right),
\end{equation}
\textcolor{black}{where $\text{Tr}({{\mathbf {G}}^{T}}\mathbf {T}{{\mathbf {G}}^{*}})={\sigma_r^2}\sum_{m=1}^{M}{\text{INR}_m}$,} and $L$ is the radar pulse length. \textcolor{black}{It is established in \cite{1998dec} that $P_D$ monotonically increases with respect to $\rho$.} Moreover, according to (\ref{detec_prob})  and (\ref{detec_prob2}), it is observed that $\text{INR}_m$ and $\mathbf{R}_x$ significantly affect the non-centrality parameter $\rho$, and are therefore required to be constrained. To generate $\mathbf{R}_x$ with good detection performance, we first generate the covariance matrix $\mathbf{R}_d$ of the ideal radar beampattern  according to \cite{wangs2022}. Under this setup, the BS designs the communication beamforming vector for optimizing the secrecy rate by exploiting the (estimated) channel difference between between Bob and Eve, and at the same time the radar  adjusts the radar waveform to jam the eavesdropper, thereby further enhancing the system secrecy rate.

\section{Joint communication beamformer and radar waveform design}

In this section, we aim to maximize the worst-case secrecy rate in (\ref{secrecycapacity}) by jointly designing radar waveform and communication beamforming vector under imperfect CSI. The optimization problem is formulated as
\begin{align} \small\label{probb2}
&\max \limits _{\substack{\mathbf{T},{\mathbf {R}_x}}} ~C_s \\
&~~\text{s.t}.~~\Vert \mathbf{R}_{x}-\mathbf{R}_{d}\Vert^{2}\leq\gamma_{p}, \tag{\ref{probb2}a}\\
&\qquad~~{\mathbf {R}_x} \succeq 0, \text{Tr} (\mathbf{R}_x) \le P_R, \tag{\ref{probb2}b}\\
&\qquad~~\mathbf{T}\succeq 0, \text{Tr} (\mathbf{T}) \le P_C, \tag{\ref{probb2}c} \\
&\qquad~~\text{rank} (\mathbf{T})=1, \tag{\ref{probb2}d} \\
&\qquad~~\text{INR}_m \le \Gamma_m, \forall m, \tag{\ref{probb2}e}
\end{align}
where $P_C$ and $P_R$ denotes the BS's and radar's transmission power budget, respectively, $\mathbf {R}_d$ is the ideal radar covariance matrix. $\gamma_{p}$ denotes the prescribed mismatch threshold between the designed covariance matrix $\mathbf {R}_x$ and $\mathbf {R}_d$. The constraint (\ref{probb2}e) limits the INR from BS to radar, and $\Gamma_m$ represents the highest toleratable INR at the $m$-th antenna of the radar.

Herein, we intend to solve the problem (\ref{probb2}) under the imperfect CSI model, which is challenging due to  the non-convex secrecy rate expression (\ref{secrecycapacity}) and the infinite number of constraints caused by the norm-bounded CSI errors. To tackle these issues, a two-layer robust cooperative design is proposed to effectively handle all possible errors, ensuring that the co-existing system can meet the performance constraints of both radar and secure communication even in worst-case scenarios.

Firstly, we employ the SDR technique \cite{sdr} to remove the rank-1 constraint in ({\ref{probb2}d}). \textcolor{black}{Since $\log(x)$ is a monotonically increasing function with respect to $x$,} (\ref{probb2}) is reformulated as (\ref{robustco1}) in the following after removing the rank-one constraint:
\begin{align}\small \label{robustco1}
\max \limits _{\substack{\mathbf{T},{\mathbf {R}_x}}} ~&\min \limits _{\substack{\bm{\Delta}_{hb}\in \mathcal{H}_b,\bm{\Delta}_{he}\in \mathcal{H}_e,\\\bm{\Delta}_{fb}\in \mathcal{F}_b,\bm{\Delta}_{fe}\in \mathcal{F}_e,\bm{{\Delta}_{m}}\in \mathcal{G}}} \frac{1+\gamma^c_b}{1+\gamma^c_e} \\
&~~~~\text{s.t}.~(\ref{probb2}\text{a}), (\ref{probb2}\text{b}),(\ref{probb2}\text{c}), (\ref{probb2}\text{e}).\notag
\end{align}

Since the objective function (\ref{robustco1}) is still a non-convex function, \textcolor{black}{we reformulate problem (\ref{robustco1}) by introducing an auxiliary variable $\mu$ as}
{\small
\begin{align} \label{robustco2}
\max \limits _{\substack{\mathbf{T},{\mathbf {R}_x}}}&\min \limits _{\substack{\bm{\Delta}_{hb}\in \mathcal{H}_b,,\bm{\Delta}_{fb}\in \mathcal{F}_b}} \frac{1+\gamma^c_b}{1+\mu} \\
&~~~~\text{s.t}.~\gamma_e^c\leq \mu,\forall \bm{\Delta}_{he}\in \mathcal{H}_e,\bm{\Delta}_{fe}\in \mathcal{F}_e, \tag{\ref{robustco2}a}\\
&\qquad~~~(\ref{robustco1}\text{a}), (\ref{robustco1}\text{b}),(\ref{robustco1}\text{c}), (\ref{robustco1}\text{d}).\notag
\end{align}}
 
It is observed from (\ref{robustco2}) that, for fixed $\mu$, the objective function $\frac{1+\gamma^c_b}{1+\mu}$ increases monotonically with respect to $\gamma_b$. Therefore, problem (\ref{robustco2}) can be decomposed into a two-layer iterative optimization, where the outer layer performs a one-dimensional search over the feasible range of $\mu$. During the search process, for fixed $\mu$, the following optimization problem is solved,
{\small
	\begin{align}\label{robust1}
\gamma^c_b(\mu)=&\max \limits _{\substack{\mathbf{T},{\mathbf {R}_x}}}~\min \limits _{\substack{\bm{\Delta}_{hb}\in \mathcal{H}_b, \bm{\Delta}_{fb}\in \mathcal{F}_b}}~~\frac{\mathbf{h}_b^H \mathbf{T}\mathbf{h}_b} {\mathbf{f}^H_b\mathbf{R}_x\mathbf{f}_b + \sigma_b^2} \\
&~\text{s.t}.~~~  \frac{\mathbf{h}_e^H \mathbf{T}\mathbf{h}_e} {\mathbf{f}^H_e\mathbf{R}_x\mathbf{f}_e + \sigma_e^2} \leq {\mu} ,\forall \bm{\Delta}_{he}\in \mathcal{H}_e,\bm{\Delta}_{fe}\in \mathcal{F}_e, \tag{\ref{robust1}a}\\
&~\qquad(\ref{robustco1}\text{a}), (\ref{robustco1}\text{b}),(\ref{robustco1}\text{c}), (\ref{robustco1}\text{d}).\notag
\end{align}}

By traversing the whole feasible region of $\mu$, the following outer layer optimization problem can be solved,
\begin{align} \small\label{inter1}
\max \limits _{{\mu}}& ~~\frac{1+\gamma^c_b(\mu)}{1+\mu} \\
&\text{s.t}.~~ \mu_{min} \leq {\mu} \leq \mu_\text{max}. \tag{\ref{inter1}a}
\end{align}

First, we find the lower and upper bound $\mu_{min}$ and $\mu_{max}$ for $\mu$ for implementing the one-dimensional search. According to (\ref{robustco2}a), one can see that the lower bound of $\mu$ is $\mu_{min}=0$. Since the secrecy rate should be greater than or equal to 0, i.e., $\text{log}_2(\frac{1+\gamma_b^c}{1+\mu})\geq0$, we have $\mu\leq\gamma_b^c$. The upper bound of $\mu$ can be expressed as
\begin{align}\small
{\mu} &\leq \frac{|\mathbf{h}_b^H \mathbf{t}|^2} {|\mathbf{f}_b^H\mathbf{x}|^2 + \sigma_b^2}\leq{|(\bar{\mathbf{h}}_b^H+\bm{\Delta^H_{hb}} )\mathbf{t}|^2}\notag\\
&\overset{(a)}{\leq} (\bar{\mathbf{h}}_b^H\mathbf{t}+{\epsilon}_{hb}\mathbf{t})^2
=|\bar{\mathbf{h}}_b^H\mathbf{t}|^2+\epsilon_{hb}^2||\mathbf{t}||_2^2+2{\epsilon_{hb}}||\mathbf{t}||_2|\bar{\mathbf{h}}_b^H\mathbf{t}|\notag\\&\overset{(b)}{\leq} P_C(||\bar{\mathbf{h}}_b^H||^2+\epsilon_{hb}^2+2{\epsilon_{hb}}|\bar{\mathbf{h}}_b^H|)\notag\\
& \triangleq \mu_\text{max},
\end{align}
\textcolor{black}{where (a) and (b) are based on (\ref{erro}) and the norm triangle inequality, respectively.}

Next, we focus on solving problem (\ref{robust1}) with given $\mu \in [\mu_{min}, \mu_{max}]$. Towards this end, we can transform problem (\ref{robust1}) by introducing an auxiliary variable $\eta$ as
{\small
	\begin{align} \label{robust2}
&\max \limits _{\substack{\mathbf{T},{\mathbf {R}_x}}}~~\eta\\
&~\text{s.t}.~~~~\frac{\mathbf{h}_b^H \mathbf{T}\mathbf{h}_b} {\mathbf{f}^H_b\mathbf{R}_x\mathbf{f}_b + \sigma_b^2} \geq \eta,\forall \bm{\Delta}_{hb}\in \mathcal{H}_b,\bm{\Delta}_{fb}\in \mathcal{F}_b, \tag{\ref{robust2}a}\\
&\qquad~~~\frac{\mathbf{h}_e^H \mathbf{T}\mathbf{h}_e} {\mathbf{f}^H_e\mathbf{R}_x\mathbf{f}_e + \sigma_e^2} \leq {\mu} ,
\forall\bm{\Delta}_{he}\in \mathcal{H}_e, \bm{\Delta}_{fe}\in \mathcal{F}_e, \tag{\ref{robust2}b} \\
&\qquad~~~(\ref{robustco1}\text{a}), (\ref{robustco1}\text{b}),(\ref{robustco1}\text{c}),(\ref{robustco1}\text{d}).\notag
\end{align}}

The constraints (\ref{robust2}a) and (\ref{robust2}b) imply that the secrecy rate under imperfect CSI is indeed the difference between the minimum legitimate rate and maximum eavesdropping rate. To address the non-convex constraints (\ref{robust2}a) and (\ref{robust2}b), we employ the Charnes-Cooper \textcolor{black}{transformation}  and  define two variables as $\widehat{\mathbf{T}}=\zeta\mathbf{T}$ and $\widehat{\mathbf{R}}_x=\zeta\mathbf{R}_x$, where the auxiliary variable $\zeta$ is given by
\begin{equation}\small
\zeta=\frac{1} {\mathbf{f}^H_b\mathbf{R}_x\mathbf{f}_b + \sigma_b^2}~\textgreater~0.
\end{equation}

The problem (\ref{robust2}) is recast as
{\small
\begin{align} \label{robust3}
&\max \limits _{\substack{\widehat{\mathbf{T}},\widehat{\mathbf {R}}_x},\zeta} ~~~\eta\\
&~\text{s.t}.~~~~ \mathbf{h}_b^H \widehat{\mathbf{T}}\mathbf{h}_b \geq \eta ,\tag{\ref{robust3}a}\forall \bm{\Delta}_{hb}\in \mathcal{H}_b,\\
&\qquad~~~\mathbf{f}^H_b\widehat{\mathbf{R}}_x\mathbf{f}_b + \zeta\sigma_b^2=1,\forall \bm{\Delta}_{fb}\in \mathcal{F}_b,\tag{\ref{robust3}b}\\
&\qquad~~~{\mathbf{h}_e^H \widehat{\mathbf{T}}\mathbf{h}_e} \leq {\mu}({\mathbf{f}^H_e\widehat{\mathbf{R}}_x\mathbf{f}_e + \zeta\sigma_e^2}), \forall \bm{\Delta}_{he}\in \mathcal{H}_e,\bm{\Delta}_{fe}\in \mathcal{F}_e,\tag{\ref{robust3}c} \\
&\qquad~~~\text{Tr}(\mathbf{g}_m^*\mathbf{g}_m^T\widehat{\mathbf{T}}) \le \zeta\Gamma_m\sigma_r^2,\forall \bm{\Delta}_{m}\in \mathcal{G}_m,\forall m, \tag{\ref{robust3}d}\\
&\qquad~~~\Vert\widehat{\mathbf{R}}_{x}-\zeta\mathbf{R}_{d}\Vert^{2}\leqslant\gamma_{p}\zeta^2, \tag{\ref{robust3}e}\\
&\qquad~~~{\widehat{\mathbf{R}}_x} \succeq 0, \text{Tr} (\widehat{\mathbf{R}}_x) \le P_R\zeta, \tag{\ref{robust3}f}\\
&\qquad~~~\widehat{\mathbf{T}}\succeq 0, \text{Tr} (\widehat{\mathbf{T}}) \le P_C\zeta, \tag{\ref{robust3}g}
\end{align}}

Now, it remains to solve problem (\ref{robust3}). Due to the existence of uncertainty constraints (\ref{robust3}a)-(\ref{robust3}d), problem (\ref{robust3}) is challenging to handle. In what follows, we utilize the S-lemma \cite{2004Convex} and its generalized form \cite{huang} to transform these constraints into linear matrix inequality (LMI) constraints.

\begin{lemma} 
	(S-lemma \cite{2004Convex}) For the quadratic inequality $f_m(x), m=1, 2$,
	\begin{equation}
	f_m(\boldsymbol{x})=\boldsymbol{x}^H \boldsymbol{A}_m \boldsymbol{x}+2 Re \{\boldsymbol{b}_m^H \boldsymbol{x}\}+c_m,
	\end{equation}
	where $\boldsymbol{x} \in \mathbb{C}^{N \times 1}, \boldsymbol{A}_m=\boldsymbol{A}_m^H \in \mathbb{C}^{N \times N}, c_m\in \mathcal{R}, b_m \in \mathbb{C}^{N\times 1}$. The implication $f_1(\boldsymbol{x}) \geq 0 \Rightarrow f_2(\boldsymbol{x}) \geq 0$ holds if and only if there exists $\lambda$ such that
	\begin{equation}
	\left[\begin{array}{ll}
	\boldsymbol{A}_2 & \boldsymbol{b}_2 \\
	\boldsymbol{b}_2^H & c_2
	\end{array}\right]-\lambda\left[\begin{array}{ll}
	\boldsymbol{A}_1 & \boldsymbol{b}_1 \\
	\boldsymbol{b}_1^H & c_1
	\end{array}\right] \succeq \mathbf{0},
	\end{equation}

	\label{lem-1}
\end{lemma}
\begin{lemma} 
	(Theorem 4.2, \cite{huang}) The quadratic matrix  inequalities
\begin{equation}
\begin{aligned}
\begin{bmatrix}
&\bm{A}&\bm{B}+\bm{CX}\\
&(\bm{B}+\bm{CX})^H&\bm{E}+\bm{X}^H\bm{F}+\bm{F}^H\bm{X}+\bm{X}^H\bm{GX}
\end{bmatrix} \succeq 0,\\
~\bm{I}-\bm{X}^H\bm{JX}\succeq 0~~\text{hold},
\end{aligned}
\end{equation}
 \textcolor{black}{if and only there exists a variable $\xi$ such that}
	\begin{equation}\label{cs3}
\begin{bmatrix}
&\bm{A}&\bm{B}&\bm{C}\\
&\bm{B}^H&\bm{E}&\bm{F}^H\\
&\bm{C}&\bm{F}&\bm{G}
\end{bmatrix}- 
\xi\begin{bmatrix}
&\bm{0}&\bm{0}&\bm{0}\\
&\bm{0}&\bm{I}&\bm{0}\\
&\bm{0}&\bm{0}&-\bm{J}
\end{bmatrix}\succeq 0.
\end{equation}
	\label{lem-2}
\end{lemma}

According to the channel uncertainty model specified in (\ref{erro}), the constraint (\ref{robust3}a) can be rewritten as
\begin{equation}	\label{imp1}
\left\{\begin{array}{l}
(\bar{\mathbf{h}_b}+\bm{\Delta_{hb}})^H \widehat{\mathbf{T}}(\bar{\mathbf{h}_b}+\bm{\Delta_{hb}})-\eta \geq 0,
\\
\Vert \bm{\Delta_{hb}} \Vert^2 \leq \epsilon_{hb}^2.
\end{array}\right.
\end{equation}

When problem (\ref{robust3}) is optimal, constraint (\ref{robust3}b) can be replaced by $\mathbf{f}^H_b\widehat{\mathbf{R}}_x\mathbf{f}_b + \zeta\sigma_b^2\leq1$ \cite{Xing2015}. By applying S-Lemma, constraint (\ref{imp1}) and constraint (\ref{robust3}b) are equivalent to
\begin{equation}\small\label{cs1}
\begin{bmatrix}
&\lambda_1 \mathbf{I}+\widehat{\mathbf{T}}&\widehat{\mathbf{T}}\bar{\mathbf{h}_b}\\
&\bar{\mathbf{h}_b}^H\widehat{\mathbf{T}}&\bar{\mathbf{h}_b}^H\widehat{\mathbf{T}}\bar{\mathbf{h}_b}-\lambda_1\epsilon_{hb}^2-\eta
\end{bmatrix} \succeq 0,
\end{equation}
\begin{equation}\small\label{cs2}
\begin{bmatrix}
&\lambda_2 \mathbf{I}-\widehat{\mathbf{R}}_x&-\widehat{\mathbf{R}}_x\bar{\mathbf{f}_b}\\
&-\bar{\mathbf{f}_b}^H\widehat{\mathbf{R}}_x&1-\bar{\mathbf{f}_b}^H\widehat{\mathbf{R}}_x\bar{\mathbf{f}_b}-\lambda_2\epsilon_{fb}^2-\zeta\sigma_b^2
\end{bmatrix} \succeq 0,
\end{equation}
\noindent where $\lambda_1$ and $\lambda_2$ are introduced as slack variables. Similarly, constraint (\ref{robust3}c) can be rewritten as
\begin{equation} \footnotesize	\label{imp2} 
\left\{\begin{array}{l}
(\bar{\mathbf{h}_e}+\bm{\Delta_{he}})^H\widehat{\mathbf{T}}(\bar{\mathbf{h}_e}+\bm{\Delta_{he}})-\mu
(\bar{\mathbf{f}_e}+\bm{\Delta_{fe}})\widehat{\mathbf{R}}_x(\bar{\mathbf{f}_e}+\bm{\Delta_{fe}})   \leq{\mu}\zeta\sigma_e^2 , 
\\
\Vert \bm{\Delta_{he}} \Vert^2 \leq \epsilon_{he}^2,~~~ \Vert \bm{\Delta_{fe}} \Vert^2 \leq \epsilon_{fe}^2.
\end{array}\right.
\end{equation}

By applying lemma 1 and lemma 2, constraint (\ref{imp2}) can be transformed into the following form
\begin{equation}\small
\begin{bmatrix} \label{cs4}
&\lambda_3\mathbf{I}-\widehat{\mathbf{T}}&-\widehat{\mathbf{T}}\bar{\mathbf{h}_e}&\bm{0}\\
&-\bar{\mathbf{h}_e}^H\widehat{\mathbf{T}}&-{Z}_1+{Z}_2+{Z}_3&\mu\bar{\mathbf{f}_e}^H\widehat{\mathbf{R}}_x\\
&\bm{0}&\mu\widehat{\mathbf{R}}_x\bar{\mathbf{f}_e}&\mu{\mathbf{R}}_x+\frac{\xi}{\epsilon_{fe}^2}\mathbf{I}
\end{bmatrix}\succeq 0,\\
\end{equation}
where ${Z}_1=\bar{\mathbf{h}_e}^H\widehat{\mathbf{T}}\bar{\mathbf{h}_e}$, ${Z}_2=\mu\bar{\mathbf{f}_e}^H\widehat{\mathbf{R}}_x\bar{\mathbf{f}_e}$, ${Z}_3=\mu\zeta\sigma_e^2-\lambda_3\epsilon_{he}^2-\xi$. $\lambda_2$, $\xi\geq 0$ are slack variables. 
For constraint (\ref{robust3}d), according to the Cauchy-Schwarz inequality and the norm-bounded error model (\ref{erro}), the following inequality holds:
\begin{align}
\small
&\text{Tr}(\mathbf{g}_m^*\mathbf{g}_m^T\mathbf{T}) =\text{Tr}((\bar{\mathbf{g}}_m+\bm{\Delta_{m}} )^*(\bar{\mathbf{g}}_m+\bm{\Delta_{m}} )^T\mathbf{T}) \notag\\
&= \text{Tr}(\bar{\mathbf{g}}^*_{\bm{m}} \bar{\mathbf{g}}_m ^T\mathbf{T})+\text{Tr}((\bar{\mathbf{g}}_m ^*\bm{\Delta}_{m}^T+\bm{\Delta^*_{m}}\bar{\mathbf{g}}_{\bm{m}} ^T+\bm{\Delta^*_{m}}\bm{\Delta}^T_{\bm{m}})\mathbf{T})\notag\\
&\leq\text{Tr}(\bar{\mathbf{g}}^*_m \bar{\mathbf{g}}_m ^T\mathbf{T})+(2{\vert\vert\bar{\mathbf{g}}_m\vert\vert}\,{\vert\vert\bm{\Delta}^T_{\bm{m}}\vert\vert}+\vert\vert\bm{\Delta}^T_{\bm{m}}\vert\vert^2)\text{Tr}(\mathbf{T})\notag\\
&\leq\text{Tr}(\bar{\mathbf{g}}^*_m \bar{\mathbf{g}}_m ^T\mathbf{T})+(2\epsilon_{gm}{\vert\vert\bar{\mathbf{g}}_m\vert\vert}+\epsilon_{gm}^2)\text{Tr}(\mathbf{T}).
\end{align}

Therefore, constraint (\ref{robust3}d) can be rewritten as
\begin{align} \small\label{cs5}
&\text{Tr}(\bar{\mathbf{g}}^*_m \bar{\mathbf{g}}_m ^T\widehat{\mathbf{T}})+(2\epsilon_{gm}{\vert\vert\bar{\mathbf{g}}_m\vert\vert}+\epsilon_{gm}^2)\text{Tr}(\widehat{\mathbf{T}}) \leq \Gamma_m\zeta{\sigma_r^2}.
\end{align}

By transforming the uncertainty constraints (\ref{robust3}a)-(\ref{robust3}d) into LMI constraints, problem (\ref{robust3}) is reformulated as follows:
\begin{align} \small\label{robust4}
\max \limits _{\substack{\widehat{\mathbf{T}},\widehat{\mathbf {R}}_x,\lambda_1,\lambda_2,\\\lambda_3,\zeta,\xi}} &~~~~~\eta\\
\text{s.t}.~~~&~~~(\ref{robust3}\text{e}),(\ref{robust3}\text{f}),(\ref{robust3}\text{g}),(\ref{cs1}),(\ref{cs2}),(\ref{cs4}),(\ref{cs5})\notag.
\end{align}

Problem (\ref{robust4}) is a convex problem that can be efficiently solved by standard convex solvers, such as CVX. After obtaining the optimized values of $\zeta,\widehat{\mathbf{T}}$ and $\widehat{\mathbf{R}}_x$, the original problem (\ref{robust3}) can be solved by recovering the solution using $\mathbf{T}=\frac{\widehat{\mathbf{T}}}{\zeta},{\mathbf{R}}_x=\frac{\widehat{\mathbf{R}}_x}{\zeta}$. \textcolor{black}{Additionally, the rank-one solution $\mathbf{t}$ can be recovered from $\mathbf{T}$ via eigenvalue decomposition or Gaussian randomization \cite{sdr}}.

Finally, we summarize the overall algorithm in Algorithm \ref{algo2}. First, we perform a one-dimensional search for $\mu$ within $[0, \mu_{\max}]$. In each search iteration, $\mu$ is fixed and the optimization problem (\ref{robust4}) is solved. After this, we choose  $\mathbf{T}$ and $\mathbf{R}_x$ that maximize the objective value $\frac{1+\gamma^c_b}{1+\mu}$ as the obtained solutions for the original problem (\ref{probb2}). The complexity of the algorithm mainly depends on the number of iterations, which is $N_{search}=\frac{\mu_{\max}}{\delta}$, as well as complexity of solving problem (\ref{robust4}). Since CVX adopts the interior-point method \cite{Wang2014}, the computational complexity of  Algorithm \ref{algo2} is $\mathcal{O}(N_{search}(4N^{6}+\allowbreak 3M^{6}+10M^2N^4+\allowbreak 9M^4N^2))$
.
\begin{algorithm}
	\caption{Two-layer Robust Cooperative Design Algorithm}
	\begin{algorithmic}[1] 
		\renewcommand{\algorithmicrequire}{\textbf{Input:}}
		\renewcommand{\algorithmicensure}{\textbf{Output:}}
		\Require Desired radar covariance matrix $\mathbf{R}_d$, power budgets $P_C$, $P_R$, mismach threshold $\gamma_{p}$, highest tolerable INR $\Gamma_m$, convergence accuracy $\epsilon$, the upper bound of channel error $\epsilon_{hb},\epsilon_{he},\epsilon_{fb},\epsilon_{fe}$, and $\epsilon_G$.
		
		\State {The search range $[0, \mu_{\max}]$ is divided into $\mu(i)$ with a search interval of $\delta$, where $i$ takes values from 1 to $\frac{\mu_{\max}}{\delta}$.}
		\Repeat
		
		\State {For the fixed $\mu(i)$, solve the optimization problem (\ref{robust4}) and record the corresponding objective $\frac{1+\gamma_b}{1+\mu(i)}$.
			
		}
		\State {Update $i$ as $i = i + 1$.}

		\Until {$i=\mu_{max}/\delta$.}
		\State {
			Let $\mu^* = \text{arg}\max\limits_{i} \frac{1+\gamma^c_b}{1+\mu(i)}$. Solve the optimization problem (\ref{robust4}) to obtain the optimal values for $\widehat{\mathbf{T}}$ and $\widehat{\mathbf{R}}_x$.}
		\State {Obtain $\mathbf{t}$ by eigenvalue decomposition.}
		
		\Ensure communication beamforming vector  $\mathbf{t}$ and radar waveform covariance $\mathbf{R}_x$.
	\end{algorithmic}\label{algo2}
\end{algorithm}
\begin{figure*}[t]
	\begin{subfigure}[]
		{\centering
			\includegraphics[width=2.48in]{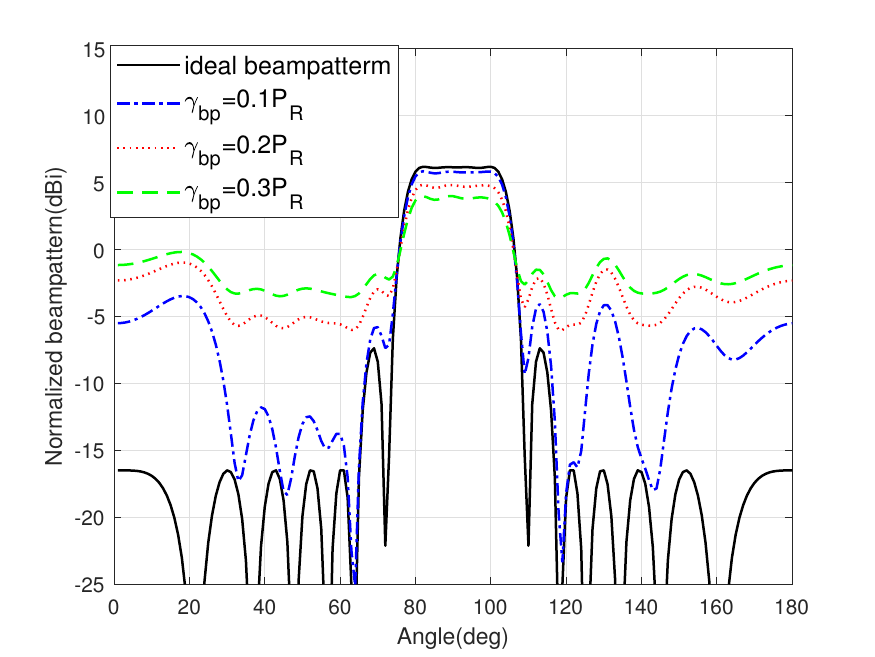}}
	\end{subfigure}
	\hspace{-7 mm}
	\begin{subfigure}[]
		{\centering
			\includegraphics[width=2.48in]{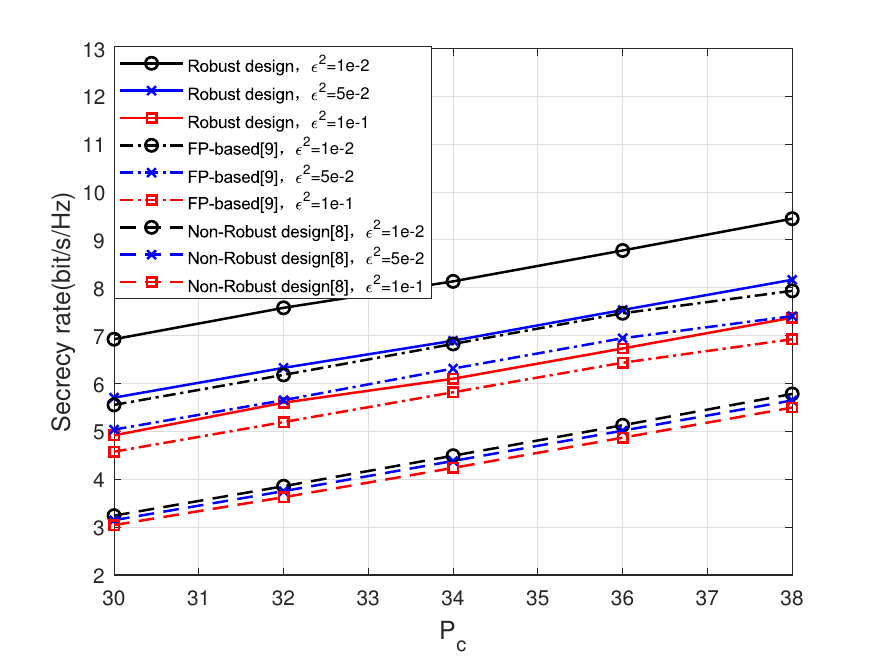}}
	\end{subfigure}
\hspace{-7 mm}
\begin{subfigure}[]
	{\centering
		\includegraphics[width=2.48in]{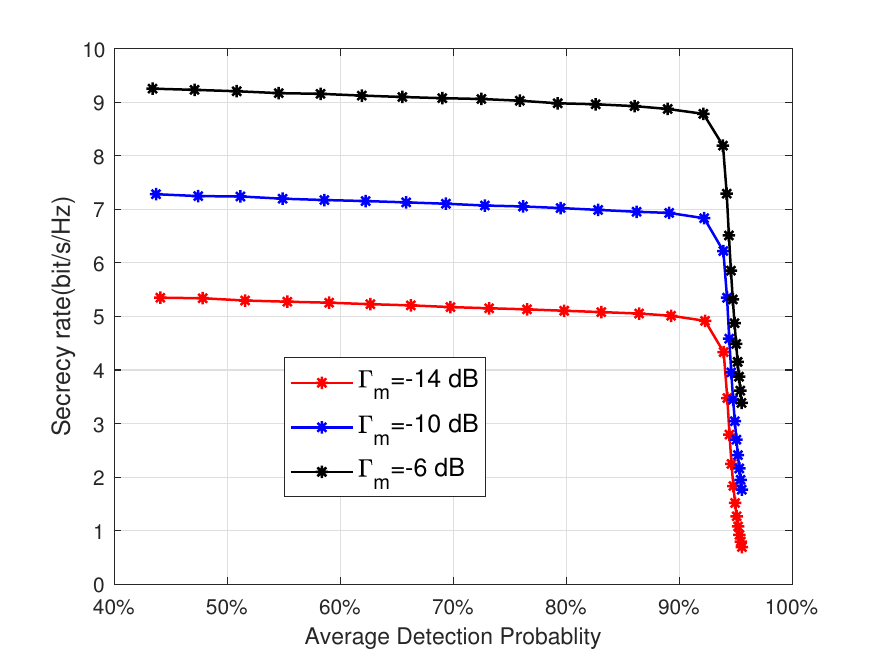}}
\end{subfigure}
	\caption{(a) Radar beampattern obtained with different mismatch threshold $\gamma_p$. (b) The secrecy rate vs. BS power $P_c$. (c) The secrecy rate vs. Average detection probability (radar SNR = 2~dB). }
\end{figure*}

\section{Simulation Results}
In this section, we show the performance of the proposed robust cooperative design algorithm via Monte Carlo simulation. Without loss of generality, we set the number of antennas at BS and radar as $N=4$ and $M=16$, respectively. During the radar detection, due to the uncertainty of the target's position, a wide beam is transmitted by the radar to detect the presence of the target. The mainlobe range of interest for the beam is set as $[80^{\circ},100^{\circ}]$. The power budget of radar and BS are set as 43 dBm and 30 dBm, respectively. The noise power is ${\sigma_b^2}={\sigma_e^2}={\sigma_r^2}={0\,\rm dBm}$. We assume that all channels follows standard Complex Gaussian distribution, i.e. $\mathbf{h}_b,\mathbf{h}_e \sim \mathcal{CN}(0, \rho_1\mathbf{I})$, $\mathbf{f}_b,\mathbf{f}_e,\mathbf{g}_m \sim \mathcal{CN}(0, \rho_2\mathbf{I})$. Since the distance from the radar to BS and Bob are typically several tens of times the distances from the BS to Bob, we set $\rho_1=1$ and $\rho_2=0.02$. Unless otherwise stated, we set the mismatch threshold to be $\gamma_{p}=0.1P_R$, the tolerable INR threshold for the radar as $\Gamma_m=-10~$dBm, radar's power budget as $P_R={43\,\rm dBm}$, and the upper bounds of channel errors as $\epsilon_{hb}^2/\rho_1 = \epsilon_{he}^2/\rho_1 = \epsilon_{fb}^2/\rho_2 = \epsilon_{fe}^2/\rho_2 = \epsilon_{gm}^2/\rho_2 = \epsilon^2 = 1 \times 10^{-2}$.

Fig.~1(a) illustrates the radar beampatterns for different mismatch thresholds $\gamma_{p}$. \textcolor{black}{The black solid line} represents the desired radar beampattern obtained as in \cite{wangs2022}. It is observed that when $\gamma_{p}=0.1P_R$, the radar beampattern closely resembles the desired beampattern, indicating that the radar maintains good sensing performance. However, as the mismatch threshold $\gamma_{p}$ increases, the gain in the mainlobe region decreases while the gain in the sidelobes increases, indicating a severer beampattern mismatch.

Fig.~1(b) shows the average secrecy rate under different BS power $P_C$ via 1000 simulations with random channel error. We compare our proposed robust optimization scheme with the Fractional Programming (FP)-based scheme in \cite{chu2022} and the non-robust scheme in \cite{Liuiccc}. It is observed that the proposed robust design algorithm achieves higher secrecy rate compared to FP-based robust scheme and the non-robust design. This is because we adopt \textcolor{black}{the exact secrecy rate} as the optimization objective, rather than the Eve's SINR. Moreover, as channel errors increase, the secrecy rate  decreases. Note that the non-robust design is more sensitive to channel errors, as it approaches the lower bound of secrecy rate even with smaller channel errors. Fig.~1(c) illustrates the sececy rates with different detection probabilities when the radar receive SNR is 2~dB, which reveals the trade-off between secure communication and sensing. As the detection probability increases, the secure performance decreases. Furthermore, as the radar tolerable INR $\Gamma_m$ increases, the degrees of freedom in optimizing communication beamforming vector increase, leading to an improvement in the secrecy rate.
\section{Conclusion}
In this paper, the robust coexistence design for MIMO radar and MISO secure communication systems was studied. Considering the presence of channel estimation errors, we formulated an optimization problem to maximize the secrecy rate by jointly optimizing the communication beamforming vector and radar waveform covariance matrix. \textcolor{black}{We aimed to enhance the security of the communication system, subject to practical constraints on the interference from the BS to radar.} By employing the S-lemma and \textcolor{black}{its generalized form,} we proposed a two-layer robust cooperative design algorithm. Simulation results demonstrated the effectiveness and robustness of the proposed algorithm, and \textcolor{black}{revealed the  trade-off between detection probability and secrecy rate.}

\bibliographystyle{IEEEtran}
\bibliography{biblp/bibfilelp}

\end{document}